
%
\def\unlockat{\catcode`\@=11}
\def\lockat{\catcode`\@=12}
\unlockat
\def\d@f@ult{} \newif\ifamsfonts \newif\ifafour
%
%
%

\font\twelverm=cmr12
\font\ninerm=cmr9
\font\sixrm=cmr6
\font\fourteenbf=cmbx12 scaled\magstep1
\font\twelvebf=cmbx12
\font\ninebf=cmbx9
\font\sixbf=cmbx6
\font\fourteeni=cmmi12 scaled\magstep1      \skewchar\fourteeni='177
\font\twelvei=cmmi12                        \skewchar\twelvei='177
\font\ninei=cmmi9                           \skewchar\ninei='177
\font\sixi=cmmi6                            \skewchar\sixi='177
\font\fourteensy=cmsy10 scaled\magstep2     \skewchar\fourteensy='60
\font\twelvesy=cmsy10 scaled\magstep1       \skewchar\twelvesy='60
\font\ninesy=cmsy9                          \skewchar\ninesy='60
\font\sixsy=cmsy6                           \skewchar\sixsy='60
\font\fourteenex=cmex10 scaled\magstep2
\font\twelveex=cmex10 scaled\magstep1

\font\ninex=cmex9

\font\sixex=cmex7 at 6pt

\font\fourteensl=cmsl10 scaled\magstep2
\font\twelvesl=cmsl10 scaled\magstep1

\font\sevensl=cmsl10 at 7pt
\font\sixsl=cmsl10 at 6pt

\font\fourteenit=cmti12 scaled\magstep1
\font\twelveit=cmti12

\font\fourteentt=cmtt12 scaled\magstep1
\font\twelvett=cmtt12
\font\fourteencp=cmcsc10 scaled\magstep2
\font\twelvecp=cmcsc10 scaled\magstep1

\newfam\cpfam
\font\fourteenss=cmss12 scaled\magstep1
\font\twelvess=cmss12
\font\tenss=cmss10
\font\niness=cmss9

\font\sevenss=cmss8 at 7pt
\font\sixss=cmss8 at 6pt
\newfam\ssfam
\newfam\msafam \newfam\msbfam \newfam\eufam
\ifamsfonts
 \font\fourteenmsa=msam10 scaled\magstep2
 \font\twelvemsa=msam10 scaled\magstep1
 \font\tenmsa=msam10
 \font\ninemsa=msam9
 \font\sevenmsa=msam7
 \font\sixmsa=msam6
 \font\fourteenmsb=msbm10 scaled\magstep2
 \font\twelvemsb=msbm10 scaled\magstep1
 \font\tenmsb=msbm10
 \font\ninemsb=msbm9
 \font\sevenmsb=msbm7
 \font\sixmsb=msbm6
 \font\fourteeneu=eufm10 scaled\magstep2
 \font\twelveeu=eufm10 scaled\magstep1
 \font\teneu=eufm10
 \font\nineeu=eufm9
 
 \font\seveneu=eufm7
 \font\sixeu=eufm6
 \def\hexnumber@#1{\ifnum#1<10 \number#1\else
  \ifnum#1=10 A\else\ifnum#1=11 B\else\ifnum#1=12 C\else
  \ifnum#1=13 D\else\ifnum#1=14 E\else\ifnum#1=15 F\fi\fi\fi\fi\fi\fi\fi}
 \def\hexmsa{\hexnumber@\msafam}
 \def\hexmsb{\hexnumber@\msbfam} 
\fi
\newdimen\b@gheight             \b@gheight=12pt
\newcount\f@ntkey               \f@ntkey=0
\def\f@m{\afterassignment\samef@nt\f@ntkey=}
\def\samef@nt{\fam=\f@ntkey \the\textfont\f@ntkey\relax}
\def\rm{\f@m0 }
\def\mit{\f@m1 }
\def\cal{\f@m2 }
\def\it{\f@m\itfam}
\def\sl{\f@m\slfam}
\def\bf{\f@m\bffam}
\def\tt{\f@m\ttfam}
\def\caps{\f@m\cpfam}
\def\ssf{\f@m\ssfam}
\ifamsfonts
 \def\msa{\f@m\msafam}  \let\bb=\msa
 \def\msb{\f@m\msbfam}
 \def\eu{\f@m\eufam}
\else
 \let \bb=\bf \let\eu=\bf
\fi
\def\fourteenpoint{\relax
    \textfont0=\fourteencp          \scriptfont0=\tenrm
      \scriptscriptfont0=\sevenrm
    \textfont1=\fourteeni           \scriptfont1=\teni
      \scriptscriptfont1=\seveni
    \textfont2=\fourteensy          \scriptfont2=\tensy
      \scriptscriptfont2=\sevensy
    \textfont3=\fourteenex          \scriptfont3=\twelveex
      \scriptscriptfont3=\tenex
    \textfont\itfam=\fourteenit     \scriptfont\itfam=\tenit
    \textfont\slfam=\fourteensl     \scriptfont\slfam=\tensl
      \scriptscriptfont\slfam=\sevensl
    \textfont\bffam=\fourteenbf     \scriptfont\bffam=\tenbf
      \scriptscriptfont\bffam=\sevenbf
    \textfont\ttfam=\fourteentt
    \textfont\cpfam=\fourteencp
    \textfont\ssfam=\fourteenss     \scriptfont\ssfam=\tenss
      \scriptscriptfont\ssfam=\sevenss
    \ifamsfonts
       \textfont\msafam=\fourteenmsa     \scriptfont\msafam=\tenmsa
         \scriptscriptfont\msafam=\sevenmsa
       \textfont\msbfam=\fourteenmsb     \scriptfont\msbfam=\tenmsb
         \scriptscriptfont\msbfam=\sevenmsb
       \textfont\eufam=\fourteeneu     \scriptfont\eufam=\teneu
         \scriptscriptfont\eufam=\seveneu \fi
    \samef@nt
    \b@gheight=14pt
    \setbox\strutbox=\hbox{\vrule height 0.85\b@gheight
                                depth 0.35\b@gheight width\z@ }}
\def\twelvepoint{\relax
    \textfont0=\twelverm          \scriptfont0=\ninerm
      \scriptscriptfont0=\sixrm
    \textfont1=\twelvei           \scriptfont1=\ninei
      \scriptscriptfont1=\sixi
    \textfont2=\twelvesy           \scriptfont2=\ninesy
      \scriptscriptfont2=\sixsy
    \textfont3=\twelveex          \scriptfont3=\ninex
      \scriptscriptfont3=\sixex
    \textfont\itfam=\twelveit    
    \textfont\slfam=\twelvesl    
      \scriptscriptfont\slfam=\sixsl
    \textfont\bffam=\twelvebf     \scriptfont\bffam=\ninebf
      \scriptscriptfont\bffam=\sixbf
    \textfont\ttfam=\twelvett
    \textfont\cpfam=\twelvecp
    \textfont\ssfam=\twelvess     \scriptfont\ssfam=\niness
      \scriptscriptfont\ssfam=\sixss
    \ifamsfonts
       \textfont\msafam=\twelvemsa     \scriptfont\msafam=\ninemsa
         \scriptscriptfont\msafam=\sixmsa
       \textfont\msbfam=\twelvemsb     \scriptfont\msbfam=\ninemsb
         \scriptscriptfont\msbfam=\sixmsb
       \textfont\eufam=\twelveeu     \scriptfont\eufam=\nineeu
         \scriptscriptfont\eufam=\sixeu \fi
    \samef@nt
    \b@gheight=12pt
    \setbox\strutbox=\hbox{\vrule height 0.85\b@gheight
                                depth 0.35\b@gheight width\z@ }}
\twelvepoint
%
%
\baselineskip = 15pt plus 0.2pt minus 0.1pt 
\lineskip = 1.5pt plus 0.1pt minus 0.1pt
\lineskiplimit = 1.5pt
\parskip = 6pt plus 2pt minus 1pt
\interlinepenalty=50
\interfootnotelinepenalty=5000
\predisplaypenalty=9000
\postdisplaypenalty=500
\hfuzz=1pt
\vfuzz=0.2pt
\dimen\footins=24 truecm 
\ifafour
 \hsize=16cm \vsize=22cm
\else
 \hsize=6.5in \vsize=9in
\fi
%
%
\skip\footins=\medskipamount
\newcount\fnotenumber
\def\clearfnotenumber{\fnotenumber=0} \clearfnotenumber
\def\fnote{\global\advance\fnotenumber by1 \generatefootsymbol
 \footnote{$^{\footsymbol}$}}
\def\fd@f#1 {\xdef\footsymbol{\mathchar"#1 }}
\def\generatefootsymbol{\iffrontpage\ifcase\fnotenumber
\or \fd@f 279 \or \fd@f 27A \or \fd@f 278 \or \fd@f 27B
\else  \fd@f 13F \fi
\else\xdef\footsymbol{\the\fnotenumber}\fi}
%
%
\newcount\secnumber \newcount\appnumber
\def\clearappnumber{\appnumber=64} \def\clearsecnumber{\secnumber=0}
\clearsecnumber \clearappnumber
\newif\ifs@c 
\newif\ifs@cd 
\s@cdtrue 
\def\unsectioned{\s@cdfalse\let\section=\subsection}
\newskip\sectionskip         \sectionskip=\medskipamount
\newskip\headskip            \headskip=8pt plus 3pt minus 3pt
\newdimen\sectionminspace    \sectionminspace=10pc
\def\Titlestyle#1{\par\begingroup \interlinepenalty=9999
     \leftskip=0.02\hsize plus 0.23\hsize minus 0.02\hsize
     \rightskip=\leftskip \parfillskip=0pt
     \advance\baselineskip by 0.5\baselineskip
     \hyphenpenalty=9000 \exhyphenpenalty=9000
     \tolerance=9999 \pretolerance=9000
     \spaceskip=0.333em \xspaceskip=0.5em
     \fourteenpoint
  \noindent #1\par\endgroup }
\def\titlestyle#1{\par\begingroup \interlinepenalty=9999
     \leftskip=0.02\hsize plus 0.23\hsize minus 0.02\hsize
     \rightskip=\leftskip \parfillskip=0pt
     \hyphenpenalty=9000 \exhyphenpenalty=9000
     \tolerance=9999 \pretolerance=9000
     \spaceskip=0.333em \xspaceskip=0.5em
     \fourteenpoint
   \noindent #1\par\endgroup }
\def\spacecheck#1{\dimen@=\pagegoal\advance\dimen@ by -\pagetotal
   \ifdim\dimen@<#1 \ifdim\dimen@>0pt \vfil\break \fi\fi}
\def\section#1{\cleareqnumber \s@ctrue \global\advance\secnumber by1
   \par \ifnum\the\lastpenalty=30000\else
   \penalty-200\vskip\sectionskip \spacecheck\sectionminspace\fi
   \noindent {\caps\enspace\S\the\secnumber\quad #1}\par
   \nobreak\vskip\headskip \penalty 30000 }
\def\undertext#1{\vtop{\hbox{#1}\kern 1pt \hrule}}
\def\subsection#1{\par
   \ifnum\the\lastpenalty=30000\else \penalty-100\smallskip
   \spacecheck\sectionminspace\fi
   \noindent\undertext{#1}\enspace \vadjust{\penalty5000}}

\def\appendix#1{\cleareqnumber \s@cfalse \global\advance\appnumber by1
   \par \ifnum\the\lastpenalty=30000\else
   \penalty-200\vskip\sectionskip \spacecheck\sectionminspace\fi
   \noindent {\caps\enspace Appendix \char\the\appnumber\quad #1}\par
   \nobreak\vskip\headskip \penalty 30000 }
\def\ack{\par\penalty-100\medskip \spacecheck\sectionminspace
   \line{\fourteencp\hfil ACKNOWLEDGEMENTS\hfil}%
\nobreak\vskip\headskip }
\def\refs{\begingroup \par\penalty-100\medskip \spacecheck\sectionminspace
   \line{\fourteencp\hfil REFERENCES\hfil}%
\nobreak\vskip\headskip \frenchspacing }
\def\endrefs{\par\endgroup}
%
%
\newif\iffrontpage \frontpagefalse
\headline={\hfil}
\footline={\iffrontpage\hfil\else \hss\twelverm
-- \folio\ --\hss \fi }
%
%
\newskip\frontpageskip \frontpageskip=12pt plus .5fil minus 2pt
\def\titlepage{\global\frontpagetrue\hrule height\z@ \relax
               \pubblock\relax }
\def\endtitlepage{\vfil\break\clearfnotenumber\frontpagefalse}
\def\title#1{\vskip\frontpageskip\Titlestyle{\caps #1}\vskip3\headskip}
\def\author#1{\vskip.5\frontpageskip\titlestyle{\caps #1}\nobreak}
\def\and{\par\kern 5pt \centerline{\sl and}}
\def\andauthor{\vskip.5\frontpageskip\centerline{and}\author}

\def\address#1{\par\kern 5pt\titlestyle{\it #1}}
\def\andaddress{\par\kern 5pt \centerline{\sl and} \address}

\def\abstract#1{\par\dimen@=\prevdepth \hrule height\z@ \prevdepth=\dimen@
   \vskip\frontpageskip\spacecheck\sectionminspace
   \centerline{\fourteencp ABSTRACT}\vskip\headskip
   {\noindent #1}}

\def\email#1{\fnote{\tentt e-mail: #1}}

%
%
\def\KUL{\address{%
   Instituut voor Theoretische Fysica, Universiteit Leuven\break
   Celestijnenlaan 200 D, B--3001 Heverlee, BELGIUM}}
\def\Bonn{\address{%
   Physikalisches Institut der Universit\"at Bonn\break
  Nu{\ss}allee 12, W--5300 Bonn 1, GERMANY}}
%

%

%
%
\newcount\refnumber \def\clearrefnumber{\refnumber=0}  \clearrefnumber
\newwrite\R@fs                              
\immediate\openout\R@fs=\jobname.refs 
\def\closerefs{\immediate\closeout\R@fs} 
\def\refsout{\closerefs\refs
\unlockat
\input\jobname.refs
\lockat
\endrefs}
\def\refitem#1{\item{{\bf #1}}}
\def\ifundefined#1{\expandafter\ifx\csname#1\endcsname\relax}
\def\[#1]{\ifundefined{#1R@FNO}%
\global\advance\refnumber by1%
\expandafter\xdef\csname#1R@FNO\endcsname{[\the\refnumber]}%
\immediate\write\R@fs{\noexpand\refitem{\csname#1R@FNO\endcsname}%
\noexpand\csname#1R@F\endcsname}\fi{\bf \csname#1R@FNO\endcsname}}
\def\refdef[#1]#2{\expandafter\gdef\csname#1R@F\endcsname{{#2}}}
%
%
\newcount\eqnumber \def\cleareqnumber{\eqnumber=0}
\newif\ifal@gn \al@gnfalse  
\def\veqnalign#1{\al@gntrue \vbox{\eqalignno{#1}} \al@gnfalse}
\def\eqnalign#1{\al@gntrue \eqalignno{#1} \al@gnfalse}
\def\(#1){\relax%
\ifundefined{#1@Q}
 \global\advance\eqnumber by1
 \ifs@cd
  \ifs@c
   \expandafter\xdef\csname#1@Q\endcsname{{%
\noexpand\rm(\the\secnumber .\the\eqnumber)}}
  \else
   \expandafter\xdef\csname#1@Q\endcsname{{%
\noexpand\rm(\char\the\appnumber .\the\eqnumber)}}
  \fi
 \else
  \expandafter\xdef\csname#1@Q\endcsname{{\noexpand\rm(\the\eqnumber)}}
 \fi
 \ifal@gn
    & \csname#1@Q\endcsname
 \else
    \eqno \csname#1@Q\endcsname
 \fi
\else%
\csname#1@Q\endcsname\fi\global\let\@Q=\relax}
%
%
\newif\ifm@thstyle \m@thstylefalse
\def\mathstyle{\m@thstyletrue}
\def\proclaim#1#2\par{\smallbreak\begingroup
\advance\baselineskip by -0.25\baselineskip%
\advance\belowdisplayskip by -0.35\belowdisplayskip%
\advance\abovedisplayskip by -0.35\abovedisplayskip%
    \noindent{\caps#1.\enspace}{#2}\par\endgroup%
\smallbreak}
\def\m@kem@th<#1>#2#3{%
\ifm@thstyle \global\advance\eqnumber by1
 \ifs@cd
  \ifs@c
   \expandafter\xdef\csname#1\endcsname{{%
\noexpand #2\ \the\secnumber .\the\eqnumber}}
  \else
   \expandafter\xdef\csname#1\endcsname{{%
\noexpand #2\ \char\the\appnumber .\the\eqnumber}}
  \fi
 \else
  \expandafter\xdef\csname#1\endcsname{{\noexpand #2\ \the\eqnumber}}
 \fi
 \proclaim{\csname#1\endcsname}{#3}
\else
 \proclaim{#2}{#3}
\fi}
\def\Thm<#1>#2{\m@kem@th<#1M@TH>{Theorem}{\sl#2}}
\def\Prop<#1>#2{\m@kem@th<#1M@TH>{Proposition}{\sl#2}}
\def\Def<#1>#2{\m@kem@th<#1M@TH>{Definition}{\rm#2}}
\def\Lem<#1>#2{\m@kem@th<#1M@TH>{Lemma}{\sl#2}}
\def\Cor<#1>#2{\m@kem@th<#1M@TH>{Corollary}{\sl#2}}
\def\Conj<#1>#2{\m@kem@th<#1M@TH>{Conjecture}{\sl#2}}
\def\Rmk<#1>#2{\m@kem@th<#1M@TH>{Remark}{\rm#2}}
\def\Exm<#1>#2{\m@kem@th<#1M@TH>{Example}{\rm#2}}
\def\Qry<#1>#2{\m@kem@th<#1M@TH>{Query}{\it#2}}
%
%
\let\Pf=\Proof

\def\<#1>{\csname#1M@TH\endcsname}
%
%
\def\ref#1{{\bf [#1]}}
\def\eg{{\it e.g.\/}}
\def\th{{\rm th}}
\def\nl{\hfil\break}
%
%
\def\qed{\vrule width 0.7em height 0.6em depth 0.2em}
\def\QED{\enspace\qed}
\def\lapprox{\hbox{\lower3pt\hbox{$\buildrel<\over\sim$}}}
\def\gapprox{\hbox{\lower3pt\hbox{$\buildrel<\over\sim$}}}
\def\quotient#1#2{#1/\lower0pt\hbox{${#2}$}}
\ifamsfonts
 \mathchardef\empty="0\hexmsb3F 
 \mathchardef\lsemidir="2\hexmsb6E 
 \mathchardef\rsemidir="2\hexmsb6F 
\else
 \let\empty=\emptyset
 \def\lsemidir{\mathbin{\hbox{\hskip2pt\vrule height 5.7pt depth -.3pt
    width .25pt\hskip-2pt$\times$}}}
 \def\rsemidir{\mathbin{\hbox{$\times$\hskip-2pt\vrule height 5.7pt
    depth -.3pt width .25pt\hskip2pt}}}
\fi
%
\def\to{\rightarrow}
\def\tto{\longrightarrow}
\def\lra{\leftrightarrow}
\def\mapright#1{\smash{
    \mathop{\tto}\limits^{#1}}}

\def\mapdown#1{\Big\downarrow
  \rlap{$\vcenter{\hbox{$\scriptstyle#1$}}$}}

\def\commdiag#1{
\def\normalbaselines{\baselineskip20pt \lineskip3pt \lineskiplimit3pt }
\matrix{#1}} 
%
\def\reals{\mathord{\bb R}} 
\def\comps{\mathord{\bb C}} 
\def\integ{\mathord{\bb Z}} 
\def\rats{\mathord{\bb Q}} 
%
\def\Tr{\mathop{\rm Tr}}
\def\ker{\mathop{\rm ker}}
\def\im{\mathop{\rm im}}
\def\underrightarrow#1{\vtop{\ialign{##\crcr
      $\hfil\displaystyle{#1}\hfil$\crcr
      \noalign{\kern-\p@\nointerlineskip}
      \rightarrowfill\crcr}}} 
\def\underleftarrow#1{\vtop{\ialign{##\crcr
      $\hfil\displaystyle{#1}\hfil$\crcr
      \noalign{\kern-\p@\nointerlineskip}
      \leftarrowfill\crcr}}}  

\def\comm#1#2{\left[#1\, ,\,#2\right]}
%
\def\pder#1#2{{{\partial #1}\over{\partial #2}}}
\def\der#1#2{{{d #1}\over {d #2}}}
%
%

\def\NPB#1#2#3{{\sl Nucl. Phys.} {\bf B#1} (#2) #3}

\def\PLB#1#2#3{{\sl Phys. Lett.} {\bf #1B} (#2) #3}
\def\JMP#1#2#3{{\sl J. Math. Phys.} {\bf #1} (#2) #3}

\def\FAaIA#1#2#3{{\sl Functional Analysis and Its Application} {\bf #1} (#2)
#3}

\def\Invm#1#2#3{{\sl Invent. math.} {\bf #1} (#2) #3}

\def\JSM#1#2#3{{\sl J. Soviet Math.} {\bf #1} (#2) #3}

\def\JETPL#1#2#3{{\sl  Sov. Phys. JETP Lett.} {\bf #1} (#2) #3}

\lockat
%
%
\def\pdo{{\hbox{\ssf $\Psi$DO}}}
\def\qpdo{{\hbox{\ssf $q$-$\Psi$DO}}}
\def\W{\mathord{\ssf W}}

\def\dop{\mathord{\ssf DOP}}

\def\Wkp{\W_{\rm KP}}
\def\Wkpq{\W_{\rm KP}^{(q)}}
\let\pb=\anticomm

\let\d=\partial
\def\kpder#1#2#3{{{\partial^{#1} #2}\over{\partial #3^{#1}}}}
\def\ad{\mathop{\rm ad}}
\def\id{\mathop{\rm id}}
\def\res{\mathop{\rm res}}
\def\comb[#1/#2]{\left[{#1\atop#2}\right]}
\def\fr#1/#2{\mathord{\hbox{${#1}\over{#2}$}}} 
\def\pair#1#2{\langle #1,#2\rangle} 
\def\mod{\mathop{\rm mod}}
\def\im{\mathop{\rm Im}}

\def\longequal{\buildrel\hbox to\wd1{\hrulefill}\over{\hbox to
\wd1{\hrulefill}}}

\def\dlb#1#2{\lbrack\!\lbrack#1,#2\rbrack\!\rbrack}

\let\isom=\cong
\def\Lie#1{\mathord{{\rm Lie}(#1)}}
\refdef[WReview]{P.~Bouwknegt and K.~Schoutens, {\sl ${\cal
W}$-Symmetry in Conformal Field Theory},  {\tt hep-th/9210010}, to
appear in {\it Phys. Reps.}}
\refdef[Univ]{J.~M.~Figueroa-O'Farrill and E.~Ramos,
\JMP{33}{1992}{833}.}
\refdef[Dickey]{L.~A.~Dickey,  {\sl Soliton equations and Hamiltonian
systems}, Advanced Series in Mathematical Physics Vol.12,  World
Scientific Publ.~Co..}
\refdef[Adler]{M.~Adler, \Invm{50}{1979}{403}.}
\refdef[WKP]{L.~A.~Dickey, {\sl Annals NY Acad.~Sci.} {\bf 491}(1987)
131;\nl J.~M.~Figueroa-O'Farrill, J.~Mas, and E.~Ramos,
\PLB{266}{1991}{298};\nl F.~Yu and Y.-S.~Wu, \NPB{373}{1992}{713}.}
\refdef[WinftyKP]{K.~Yamagishi, \PLB{259}{1991}{436};\nl F.~Yu and
Y.-S.~Wu, \PLB{236}{1991}{220}}
\refdef[Class]{J.~M.~Figueroa-O'Farrill and E.~Ramos, \PLB{282}{1992}{357}
({\tt hep-th/9202040}); {\sl Classical ${\cal W}$-algebras from
dispersionless Lax hierarchies}, preprint to appear.}
\refdef[Radul]{A.~O.~Radul, \JETPL{50}{1989}{371},
\FAaIA{25}{1991}{25};\nl
I.~Vaysburd and A.~Radul, \PLB{274}{1992}{317}.}
\refdef[WKPq]{J.~M.~Figueroa-O'Farrill, J.~Mas, and E.~Ramos, {\sl A
One-Parameter Family of Hamiltonian Structures for the KP Hierarchy
and a Continuous Deformation of the Nonlinear $\W_{\rm KP}$ Algebra},
Preprint BONN--HE--92--20, US--FT--92/7, KUL--TF--92/20,
{\tt hep-th/9207092}.}
\refdef[Winftys]{J.~M.~Figueroa-O'Farrill, J.~Mas, and E.~Ramos, {\sl
The Topography of $\W_\infty$-Type Algebras}, {\tt hep-th/9208077},
to appear in {\sl Phys. Lett. B}.}
\refdef[Wgeom]{J.~M.~Figueroa-O'Farrill, E.~Ramos, and S.~Stanciu,
{\sl A Geometrical Interpretation of Classical $\W$-Transformations},
{\tt hep-th/9209002}, to appear in {\sl Phys. Lett. B}.}
\refdef[FF]{B.~Feigin and E.~Frenkel, {\sl Affine Kac--Moody Algebras
at the Critical Level and Gelfand--Dikii Algebras}, in the Proceedings
of the RIMS Research Project 1991, {\it Infinite Analysis}.}
\refdef[AddSym]{S.~Aoyama and Y.~Kodama, \PLB{278}{1992}{56}.}
\refdef[KW]{B.~A.~Kupershmidt and G.~Wilson, \Invm{62}{1981}{403}.}
\refdef[KZ]{B.~Khesin and I.~Zakharevich, {\sl Lie-Poisson group of
pseudodifferential symbols and fractional KP-KdV hierarchies},
IHES/MIT preprint, to appear in {\sl C.R. Acad. Sci.}.}
\refdef[DS]{V.~G.~Drinfel'd and V.~V.~Sokolov, \JSM{30}{1984}{1975}.}
\refdef[KK]{B.~A.~Khesin and O.~S.~Kravchenko, \FAaIA{25}{1991}{152}.}
\refdef[DickeyAS]{L.~A.~Dickey, {\sl Additional Symmetries of KP,
Grassmannian, and the String Equation}, {\tt hep-th/9204092}; {\sl
Part II} {\tt hep-th/9210155}.}
\refdef[DSS]{A.~Das, E.~Sezgin, and S.~J.~Sin, \PLB{277}{1992}{435}.}
\overfullrule=0pt
\mathstyle
\def\pubblock{ \line{\hfil\rm BONN--HE--92--34}
               \line{\hfil\rm KUL--TF--92/27}
               \line{\hfil\rm QMW--PH--92--20}
               \line{\hfil\tt hep-th/9211036}
               \line{\hfil\rm November 1992}}
\titlepage
\title{The Algebra of Differential Operators on the Circle and
$\Wkpq$}
\author{Jos\'e~M.~Figueroa-O'Farrill\email{jmf@avzw01.physik.uni-bonn.de}}
\Bonn
\andauthor{Eduardo Ramos\email{ramos@v1.ph.qmw.ac.uk}}
\KUL
\andaddress{Department of Physics, Queen Mary and Westfield
College\break
Mile End Road, London E1 4NS, UK\fnote{\tenrm Present address.\hfil}}
\abstract{Radul has recently introduced a map from the Lie algebra of
differential operators on the circle to $\W_n$.  In this note we
extend this map to $\Wkpq$, a recently introduced one-parameter
deformation of $\Wkp$---the second hamiltonian structure of the KP
hierarchy.  We use this to give a short proof that $\W_\infty$ is the
algebra of additional symmetries of the KP equation.}
\endtitlepage

\section{Introduction}

$\W$-algebras (see \[WReview] for a timely review) are one of the most
interesting algebraic structures to have appeared in theoretical
physics in recent times.  Besides the fundamental role
they play in physics---both in the theory of integrable systems and in
two-dimensional conformal field theory---$\W$-algebras are beginning
to show up also in mathematics, for example, in Drinfel'd's approach
to the generalized Langlands correspondence \[FF].  Despite this
growing interest on both sides of the fence, a deep understanding of
these algebras is to a large extent lacking and fundamental questions
concerning their geometrical significance and representation theory
remain unanswered (although see \[Wgeom]).  In the absence of a
clearer path to follow, one hopes that one can make some progress by
relating $\W$-algebras to algebraic structures which are better
understood.  This drive towards determining the right place for
$\W$-algebras in the algebraic world characterizes much of current
research on this topic, and this paper is no exception.

One particularly fruitful method of investigating how a class of
algebras fits within other algebraic structures is to try and
establish maps (morphisms) between its objects and other well-known
objects.  In the case of $\W$-algebras, examples of such maps are the
Miura transformation \[KW], the (generalized) Drinfel'd--Sokolov
reduction \[DS], and the Radul map \[Radul].  In a few words, the
Miura transformation embeds the $\W$-algebra in the universal
envelope of the algebra of free fields (Heisenberg algebras); whereas
the Drinfel'd--Sokolov reduction exhibits the $\W$-algebra as a
subquotient of the universal enveloping algebra of an affine Lie
algebra.  On the other hand, the Radul map is a different kind of
morphism.  $\W$-algebras act naturally as infinitesimal canonical
transformations on the space of Lax operators and the Radul map is a
Lie algebra homomorphism from the differential operators on the
circle to the algebra of vector fields on the space of Lax operators.
Some of these vector fields generate $\W$-tranformations.

Brevity demands that we omit a discussion of these maps.  There is
ample literature already on the first two maps (see, for example,
\[WReview] and references therein) and the Radul map is leisurely
described in the second reference of \[Radul].  However, the point of
this note being to introduce a generalization of the Radul map to the
space of generalized pseudodifferential operators, we briefly review
this formalism in section 2.  Section 3 contains the main result of
this paper:  the extension of the Radul map to the space of
generalized pseudodifferential Lax operators.  Section 4 contains an
application of this extension to the case of the KP hierarchy.  Using
the homomorphism property of the generalized Radul map one can make
transparent the appearance of $\W_\infty$ as the algebra of additional
symmetries \[AddSym] of the KP hierarchy.  Finally, section 5 contains
some concluding remarks.

\section{Generalized Pseudodifferential Operators and $\Wkpq$}

In this section we briefly review the extension of the formalism of
\[Dickey] to encompass generalized pseudodifferential operators.  We
follow \[WKPq] to which the reader is referred for more details.  A
description of the underlying Lie-Poisson structures can be found in
\[KZ].

By a pseudodifferential symbol we mean a formal Laurent series in a
parameter $\xi^{-1}$ of the form $P(z,\xi) = \sum_{i=-\infty}^{\rm
finite} p_i(z)\xi^i$ whose coefficients we take to be smooth functions
on the circle.  Symbols have a commutative multiplication given by
multiplying the Laurent series; but one can also define a composition
law (denoted by $\circ$)
$$P(z,\xi)\circ Q(z,\xi)= \sum_{k\geq 0} {1\over k!}
\kpder{k}{P}{\xi} \kpder{k}{Q}{z}~,\(symbolcomp)$$
making the map
$$P(z,\xi) \mapsto \sum_i p_i(z) \d^i\(symtops)$$
from pseudodifferential symbols to pseudodifferential operators
(\pdo's) into an algebra homomorphism.  Conversely to every
\pdo\ we associate its symbol by first writing all $\d$'s to the
right and then substituting $\d$ with $\xi$.

The advantage of working with symbols is that symbol composition is
a well-defined operation on arbitrary smooth functions of $z$ and
$\xi$.  For example, for $a=a(z)$,
$$\log\xi \circ a = a\log\xi - \sum_{j=1}^\infty
{(-1)^j\over j} a^{(j)} \xi^{-j}~,\(logderiv)$$
which shows that the commutator (under symbol composition) with
$\log\xi$, denoted by $\ad\log\xi$, is an outer derivation on the
algebra of pseudodifferential symbols.  Similarly, if $q$ is any
complex number, not necessarily an integer, we find
$$\xi^q \circ a = \sum_{j=0}^\infty \comb[q/j] a^{(j)}
\xi^{q-j}~,\(qLeibniz)$$
where we have introduced, for $q$ any complex number, the generalized
binomial coefficients $\comb[q/j] \equiv [q]_j/j!$, where $[q]_j
\equiv q(q-1)\cdots(q-j+1)$ is the Pochhammer symbol.  Conjugation by
$\xi^q$ is therefore an outer automorphism of the algebra of
pseudodifferential symbols, which is the integrated version of $\ad
\log\xi$.  Hence, via their associated symbols, one can give a
well-defined meaning to objects such as the logarithm of the
derivative $\log\d$ and to an arbitrary complex power of the
derivative $\d^q$.  From now on, we shall work with these formal
objects with the tacit understanding that they are defined via their
symbols.

It follows from \(qLeibniz) that multiplication (either on the left or
on the right) by $\d^q$ sends \pdo's into operators of the form
$\sum_{j\leq N} p_j(z) \d^{q+j}$.  Let us denote the set of these
operators by ${\cal S}_q$.  It is clear that ${\cal S}_q$ is a
bimodule over the algebra of \pdo's, which for $q\in\integ$ coincides
with the algebra itself.  In fact, since ${\cal S}_q = {\cal S}_p$ for
$p \equiv q \mod \integ$, we will understand ${\cal S}_q$ from now on
as implying that $q$ is reduced modulo the integers.  Moreover, symbol
composition induces a multiplication ${\cal S}_p \times {\cal S}_q \to
{\cal S}_{p+q}$, where we add modulo the integers.  We call their
union ${\cal S}=\cup_q {\cal S}_q$, the algebra of generalized
pseudodifferential operators (\qpdo's).

On ${\cal S}_0$ one can define a trace form as follows.  Let us define
the residue of a \pdo\ $P = \sum_{j\leq N} p_j(z) \d^j$ by $\res
P = p_{-1}(z)$.  Then one defines the Adler trace \[Adler] as $\Tr P =
\int \res P$, where $\int$ is integration over the circle.  As the
name suggests, $\Tr\comm{P}{Q}=0$, since the residue of a commutator
is a total derivative.  The Adler trace can be used to define a
symmetric bilinear form on \pdo's
$$\pair{A}{B} \equiv \Tr A B~,\(sbf)$$
which extends to a symmetric bilinear form on all of ${\cal S}$.
Relative to this form the dual space to ${\cal S}_q$ is clearly
isomorphic to ${\cal S}_{-q}$.

The ring ${\cal S}_0$ of \pdo's splits into the direct sum of two
subrings ${\cal S}_0 = {\cal R}_+ \oplus {\cal R}_-$, corresponding to
the differential and integral operators, respectively.  This
decomposition is a maximally isotropic split for the bilinear form
\(sbf), since $\Tr A_\pm  B_\pm = 0$, where $A_\pm$ denotes the
projection of $A$ onto ${\cal R}_\pm$ along ${\cal R}_\mp$.

In order to define the generalized Adler map, we need to briefly
introduce some formal geometry on the space ${\cal M}_q$ of \qpdo's of
the form
$$L = \d^q + \sum_{j=1}^\infty u_j(z)\d^{q-j}~.\()$$
The tangent space ${\cal T}_q$ to ${\cal M}_q$ is parametrized by the
infinitesimal deformations of $L$, which are given by \qpdo's of the
form
$$ A = \sum_{j=1}^\infty a_j(z) \d^{q-j}~.\(vectors)$$

One-forms are parametrized by the dual space ${\cal T}^*_q$ of ${\cal
T}_q$ under the bilinear form \(sbf).  That is, ${\cal T}^*_q$ is made
up of \qpdo's of the form
$$X=\sum_{j=1}^\infty \d^{j-q-1} x_j~\mod \d^{-q}{\cal
R}_-~.\(oneform)$$
In other words, ${\cal T}_q^* \isom {\cal S}_{-q} / \d^{-q}{\cal
R}_-$.  For $A$ and $X$ as above,
$$\Tr A X = \sum_{j=1}^\infty \int a_j x_j~,\(pairing)$$
which is clearly nondegenerate.

The generalized Adler map is defined as follows.  If $X\in {\cal
S}_{-q}$, define
$$\veqnalign{J(X) &= (L X)_+ L - L (XL)_+\cr
&= L (XL)_- - (L X)_- L~.\(qAdler)\cr}$$
Notice that since $L X$ and $X L$ belong to ${\cal S}_0$, it makes
sense to project onto their integral and/or differential parts.
Notice also that the first equation in \(qAdler) implies that
$\d^{-q}{\cal R}_- \subset \ker J$, whereas the second equation tells
us that $\im J \subset {\cal T}_q$.  Furthermore if $Y$ is another
1-form,
$$\Tr J(X) Y = - \Tr X J(Y)~;\(skew)$$
hence $J$ induces a skewsymmetric linear map  ${\cal T}^*_q \to {\cal
T}_q$---also denoted by $J$.  We can therefore use it to define a
bracket on the functions on ${\cal M}_q$ as follows:
$$\pb{F}{G} = \Tr J(dF)  dG~,\(bracket)$$
where for $F$ a function in ${\cal M}_q$, its gradient is defined
implicitly by
$$\Tr A\,dF = \left.\der{}{\epsilon} F[L+\epsilon A]
\right|_{\epsilon=0}~,\(gradient)$$
for all tangent vectors $A\in{\cal T}_q$.

It is proven in \[WKPq] that this is indeed a Poisson bracket, and
moreover, that it defines on the coefficients $\{u_i\}$ of $L$ a
one-parameter $\W$-algebra, called $\Wkpq$, from which all $\W_n$ can
be obtained by reduction.  Furthermore, as showed in \[Winftys], all
hitherto known $\W_\infty$-type algebras can also be obtained from
$\Wkpq$ by contractions and/or reductions.

\section{The Generalized Radul Map}

Consider the subring ${\cal R}_+$ of differential operators on the
circle and give it a Lie algebra structure by the commutator.  We call
the resulting Lie algebra $\dop$.  The generalized Radul map is
defined as follows.  Fix a \qpdo\ $L\in {\cal M}_q$ and define
$W:\dop \to {\cal T}_q$ by
$$W(E) = (LEL^{-1})_-L = LE - (LEL^{-1})_+L~.\(radul)$$
On ${\cal T}_q$ we can define a Lie bracket as follows.  Every
$A\in{\cal T}_q$ of the above form defines a vector field $\d_A$
by
$$\d_A = \sum_{i=1}^\infty\sum_{k=0}^\infty a_i^{(k)}
\pder{~}{u_i^{(k)}}~.\(vecfield)$$
In particular, $\d_A L = A$.  We then define the Lie bracket
$\dlb{A}{B}$ of two vectors $A,B\in{\cal T}_q$ by
$$\d_{\dlb{A}{B}} = \comm{\d_A}{\d_B}~;\(bratang)$$
or equivalently
$$\dlb{A}{B} = \d_AB - \d_BA~.\()$$
(Notice that this is not the ordinary commutator in ${\cal S}_q$, which
would send the bracket of $A,B$ to ${\cal S}_{2q}$.)

\Thm<Mor>{If $E,F\in\dop$ are independent of $L$ (that is, their
coefficients do not depend on the coefficients of $L$) then
$$\dlb{W(E)}{W(F)}=W(\comm{E}{F})~.$$}

\Pf Because of \(bratang) this is equivalent to
$$\comm{\d_{W(E)}}{\d_{W(F)}} = \d_{W(\comm{E}{F})}~,$$
which we proceed to prove.  Acting on $L$,
$$\eqnalign{\comm{\d_{W(E)}}{\d_{W(F)}}& L\cr
={}& \d_{W(E)} W(F) - (E\lra F)\cr
={}& \d_{W(E)} (LFL^{-1})_-L - (E\lra F)\cr
={}& (L\d_{W(E)}F L^{-1})_-L + (W(E)FL^{-1})_-L\cr
&- (LFL^{-1}W(E)L^{-1})_-L + (LFL^{-1})_-W(E) - (E \lra F)\cr
={}& W(\d_{W(E)}F) + ((LEL^{-1})_-LFL^{-1})_-L\cr
& - (LFL^{-1}(LEL^{-1})_-)_-L + (LFL^{1-})_-(LEL^{-1})_-L - (E\lra
F)\cr
={}& W(\d_{W(E)}F) + ((LEL^{-1})_-LFL^{-1})_-L \cr
&+ (LEL^{-1}(LFL^{-1})_-)_-L - (LEL^{1-})_-(LFL^{-1})_-L - (E\lra
F)\cr
={}& W(\d_{W(E)}F) + ((LEL^{-1})_-(LFL^{-1})_+)_-L\cr
&+ (LEL^{-1}(LFL^{-1})_-)_-L - (E\lra F)\cr
={}& W(\d_{W(E)}F) + (LEFL^{-1})_-L - (E\lra F)~,\cr}$$
whence
$$\comm{\d_{W(E)}}{\d_{W(F)}} L = W(\d_{W(E)}F - \d_{W(F)}E +
\comm{E}{F})~.\(mor)$$
If $E$ and $F$ are independent of $L$, the variations
$\d_{W(F)}E$, $\d_{W(E)}F$ vanish and the theorem follows. \QED

\Rmk<>{It follows from \(mor) that if $E$ and $F$ depend on $L$ then
we have to correct the Lie bracket in $\dop$ to preserve the
homomorphism. Let us define the improved bracket on $\dop$ by
$$\comm{E}{F}_c \equiv \d_{W(E)}F - \d_{W(F)}E + \comm{E}{F}~.\()$$}

In \[WKPq] it is proven that the image of the generalized Adler map is
a subalgebra of ${\cal T}_q$, and this allows us to pull back the
bracket $\dlb{}{}$ on ${\cal T}_q$ to a bracket $\comm{}{}_L^*$ on
${\cal T}_q^*$ defined by requiring that the Adler map be a
homomorphism.  Explicitly, for $X,Y\in{\cal T}_q^*$,
$$\dlb{J(X)}{J(Y)} = J(\comm{X}{Y}_L^*)~,\()$$
where
$$\comm{X}{Y}_L^* = \d_{J(X)}Y + X(LY)_- - (XL)_+Y - (X \lra
Y)~.\(bracot)$$

Therefore we have two Lie algebra homomorphisms to ${\cal T}_q$:
$$
\commdiag{&&{\cal T}_q^*\cr
&&\mapdown{J}\cr
\dop&\mapright{W}& {\cal T}_q\cr}$$
and it is natural to ask whether the triangle can be completed.  It
turns out that this is indeed the case.

\Thm<Mortoo>{There exists a Lie algebra isomorphism $R:\dop \to {\cal
T}_q^*$ making the following diagram commutative
$$
\def\mapne#1{\llap{$\vcenter{\hbox{$\scriptstyle #1$}}$}\nearrow}
\commdiag{&&{\cal T}_q^*\cr
&\mapne{R\,}&\mapdown{J}\cr
\dop&\mapright{W}& {\cal T}_q\cr}$$}

\Pf Define $R(E) = -EL^{-1} \mod \d^{-q}{\cal R}_-$.  It is clear from
\(qAdler) that $\d^{-q}{\cal R}_- \subset \ker J$; whence
$$J(R(E)) = J(-EL^{-1}) = (LEL^{-1})_-L - L(EL^{-1}L)_- = W(E)~.\()$$

We now prove that it is bijective.  Consider the map $S:{\cal S}_{-q}
\to \dop$ given by $S(X)= -(XL)_+$.  Its kernel is precisely
$\d^{-q}{\cal R}_-$, whence it induces a one-to-one map $R^{-1}: {\cal
T}_q^* \to \dop$, which as the name suggests is inverse to $R$.
Indeed,
$$\eqnalign{R\circ S(X) ={}& (XL)_+L^{-1} \mod \d^{-q}{\cal R}_-\cr
={}& X - (XL)_-L^{-1} \mod \d^{-q}{\cal R}_-\cr
={}& X \mod \d^{-q}{\cal R}_-~,\cr}$$
whence $R\circ R^{-1} = \id$.  Conversely,
$$R^{-1}\circ R(E) = - S(EL^{-1}) = E_+ = E~.$$

Finally we prove that $R$ is a Lie algebra homomorphism---whence it
will follow that so is $R^{-1}$.  Notice that since $J\circ R = W$ and
both $J$ and $W$ are homomorphisms, $R$ must be a homomorphism modulo
$\ker J$.  Nevertheless a short calculation shows that the
homomorphism is exact.  First notice that $\d^{-q}{\cal R}_-$ is an
ideal relative to the extension of $\comm{}{}_L^*$ to $S_{-q}$.
Therefore,
$$\eqnalign{\comm{R(E)}{R(F)}_L^* ={}& \comm{EL^{-1}}{FL^{-1}}_L^* \cr
={}& -\d_{W(E)}(FL^{-1}) + EL^{-1}(LFL^{-1})_- - EFL^{-1} - (E\lra
F)\cr
={}& -\d_{W(E)}FL^{-1} + FL^{-1}W(E)L^{-1} + EL^{-1}W(F)L^{-1}\cr
& - EFL^{-1} - (E\lra F)\cr
={}& - \left( \d_{W(E)}F - \d_{W(F)}E + \comm{E}{F}\right) L^{-1}\cr
={}& R(\comm{E}{F}_c)~,\cr}$$
which proves the proposition. \QED

\Rmk<UrRadul>{If we take $q=N\geq 2$ and we consider the subspace
$\overline{\cal M}_N \subset {\cal M}_N$ consisting only of
differential operators, we recover the original Radul map.  In fact,
let $\overline{\cal T}_N$ denote the tangent space and $\overline{\cal
T}_N^* \isom {\cal R}_-/\d^{-N}{\cal R}_-$ denote its dual.  Then we
have Lie algebra homomorphisms
$$\eqnalign{\overline{R}:\dop &\to \overline{\cal T}_N^*\cr
           E &\mapsto -(EL^{-1})_-~\mod \d^{-N}{\cal R}_-\cr
\noalign{\hbox{and}}
            \overline{W}:\dop &\to \overline{\cal T}_N\cr
           E &\mapsto (LEL^{-1})_-L\cr}$$
making the following diagram commutative
$$
\def\mapne#1{\llap{$\vcenter{\hbox{$\scriptstyle #1$}}$}\nearrow}
\commdiag{&&\overline{\cal T}_N^*\cr
&\mapne{\overline{R}\,}&\mapdown{J}\cr
\dop&\mapright{\overline{W}}& \overline{\cal T}_N\cr}$$
Unlike in \<Mortoo>, however, $\overline{R}$ is no longer one-to-one.  In
fact, for any $F\in\dop$, $FL\in\ker\overline{R}$.}

\section{Dressing Transformations and a Radul-type Map}

In this section we discuss an immediate application of the
homomorphism property of the Radul map.  This concerns the recent
identification \[AddSym] of $\W_\infty$ as the algebra of additional
symmetries (see, for example, \[Dickey] and also \[DickeyAS]) of the
KP hierarchy.  We start with a brief review of dressing
transformations.  We follow \[Dickey].

The Volterra group $G=1 + {\cal R}_-$ acts naturally on ${\cal M}_q$
via dressing transformations $L\mapsto \Phi^{-1} L \Phi$, where
$G\ni\Phi=1+\sum_1^\infty w_i\d^{-i}$.  Since the coefficient $u_1$ of
$L$ does not evolve with the KP flows and is invariant under dressing
transformations, we will restrict ourselves to those operators with
$u_1=0$.  By undressing such an operator $L$
$$L\Phi = \Phi\d^q\(undressing)$$
we can lift the KP flows to flows on the Volterra group.  In the
$q$-formulation of the KP hierarchy (see \[WKPq]) the KP flows are
given by
$$\d_n L = \comm{L^{n/q}_+}{L} = \comm{L}{L^{n/q}_-}~,\(KPflows)$$
where $\d_n$ stands for $\pder{}{t_n}$, the derivative along the
$n^\th$ ``time'' of the KP hierarchy.

\Prop<Flind>{The KP flows \(KPflows) on $L$ are induced by the
following flows on $\Phi$:
$$\d_n \Phi = -(\Phi \d^n \Phi^{-1})_-\Phi~.\(Gflows)$$}

\Pf From \(undressing), $L=\Phi \d^q \Phi^{-1}$.  Hence,
$$\eqnalign{\d_n L ={} & \d_n\Phi \d^q \Phi^{-1} - \Phi\d^q\Phi^{-1}
\d_n\Phi \Phi^{-1}\cr
={}& - (\Phi\d^n\Phi^{-1})_- L + L (\Phi\d^n\Phi^{-1})_- \cr
={}& \comm{L}{L^{n/q}_-}~.~\QED\cr}$$

\Rmk<Powers>{In the last line of the proof we have used the fact that
if $A = \Phi B \Phi^{-1}$ then $A^z = \Phi B^z \Phi^{-1}$ for all
$z\in\comps$.  This fact follows immediately for $z\in\rats$.  For
$z\in\reals$ it follows from continuity of the map $z\mapsto P^z$ (for
all formally elliptic \pdo's $P$) and for $z\in\comps$ it is a
consequence of the fact that this map is actually holomorphic (see,
for example, \[WKPq]).}

The expression \(Gflows) is reminiscent of the Radul map \(radul).  In
fact, if we define a map $W':\dop \to \Lie{G}\isom {\cal R}_-$ by
$$W'(E) = (\Phi E \Phi^{-1})_-\Phi~,\(radultoo)$$
then $\d_n\Phi = -W'(\d^n)$.  In general, for $E\in\dop$ we define a
flow $\d_{W'(E)}$ on the Volterra group by $\d_{W'(E)}\Phi = W'(E)$.
The proof of \<Mor> carries over {\it mutatis mutandis} to a proof of
the analogous result:

\Thm<Mortoo>{For all $E,F\in\dop$
$$\comm{\d_{W'(E)}}{\d_{W'(F)}}=\d_{W'(\comm{E}{F}_c)}~,$$
where
$$\comm{E}{F}_c \equiv \d_{W'(E)} F - \d_{W'(F)} E +
\comm{E}{F}~.\(impbra)$$}

\Cor<Flowscommute>{The KP flows \(KPflows) commute.}

\Pf By \<Flind> it is enough to prove that the flows \(Gflows)
commute.  But by the theorem, $\comm{\d_n}{\d_m} =
\d_{W'(\comm{\d^n}{\d^m})} = 0$. \QED

Consider now the formal differential operator
$$\Gamma = \sum_{j=1}^\infty j t_j \d^{j-1}~,\(Gamma)$$
where $t_j$ is the $j^\th$ KP time.  It's clear that
$\comm{\d_n}{\Gamma}=n\d^{n-1}$, and it follows from \(KPflows) that
$\d_1=\d$, whence
$$\comm{\d_n-\d^n}{\Gamma}=0~.\(trivid)$$
Upon dressing, this relation becomes
$$\comm{\d_n - L^{n/q}_+}{M}=0~,\()$$
where $M\equiv \Phi\Gamma\Phi^{-1}$.  Since $\comm{\d}{\Gamma}=1$, the
Lie algebra ${\cal A}$ generated by $\Gamma^k \d^m$, for $k\geq 0$ and
$m\in\integ$ is isomorphic to $\dop$ (relative to the commutator).
And so is the algebra generated by $M^k L^{m/q}$, for $k\geq 0$ and
$m\in\integ$, since dressing transformations are algebra
automorphisms.  Explicitly, the isomorphism $\dop \to {\cal A}$ is
given by $z\mapsto -\d$ and $\d/\d z \mapsto \Gamma$.  It may seem at
first more natural to send $\d/\d z$ to $\d$ and $z$ to $\Gamma$, but
in view of the applications we have in mind, we prefer this choice.

Let us define flows on $G$ by
$$\d_{mk} \Phi = W'(\Gamma^k\d^m)~.\(Moreflows)$$
(Notice that $\d_{m0} = -\d_m$.)  As corollaries of \<Mortoo> we get
two important results concerning these flows.

\Cor<addsyms>{The flows are symmetries of the KP hierarchy; in other
words, they commute with the Lax flows \(KPflows).}

\Pf By \<Mortoo>,
$$\comm{\d_{mk}}{\d_n} = \d_{-W'(\comm{\Gamma^k\d^m}{\d^n}_c)}~;$$
where by \(impbra)
$$\eqnalign{\comm{\Gamma^k\d^m}{\d^n}_c ={}& (\d_n\Gamma^k)\d^m +
\comm{\Gamma^k\d^m}{\d^n}\cr
={}& \comm{\d_n-\d^n}{\Gamma^k\d^m}~,\cr}$$
which vanishes by \(trivid). \QED

\Rmk<>{The flows $\d_{mk}$ for $k\geq 1$, $m\in\integ$ are the
additional symmetries of the KP hierarchy (see, \eg, \[Dickey]).}

\Cor<winfty>{(Aoyama--Kodama \/\[AddSym]) The additional symmetries
generate a Lie algebra isomorphic to $\W_\infty$.}

\Pf By \<Mortoo>, the algebra of the additional flows is isomorphic to
the subalgebra of {\cal A} generated by $\Gamma^k\d^m$, for $k\geq1$
and $m\in\integ$.  But under the isomorphism $\dop \to {\cal A}$
described above, this is isomorphic to the subalgebra of $\dop$
consisting of differential operators without zeroth order piece---in
other words, $\W_\infty$. \QED

\Rmk<CenExt>{Although realized here without it, $\W_\infty$ has a
natural central extension given, as a subalgebra of $\dop$, by the
Khesin--Kravchenko \[KK] cocycle.  In the KP context, the central
extension appears when acting on the $\tau$-functions---equivalently,
when we realize $\W_\infty$ as free fermion bilinears in a
two-dimensional conformal field theory.}

\section{A Few Concluding Remarks}

In this paper we have exhibited a Lie algebra isomorphism between the
algebra $\dop$ of differential operators on the circle and a
subalgebra of the algebra of vector fields on the space of generalized
pseudodifferential operators (\qpdo's) of the form $\d^q +
\sum_{i=1}^\infty u_i(z) \d^{q-i}$.  The subalgebra in question is
given by the image of the Adler map and contains $\Wkpq$ as a
subalgebra.

Moreover, a simple and conceptual proof of the fact \[AddSym] that
$\W_\infty$ is the algebra of additional symmetries of the KP
hierarchy followed from this result.  We did this by using the
homomorphism property of a Radul-type map sending $\dop$ to flows on
the Volterra group.  This proof extends to the supersymmetric case.
For the case of the Manin--Radul SKP hierarchy, the algebra of
additional symmetries has already been identified \[DSS]; but from
the point of view of dressing transformations, it is the Jacobian SKP
hierarchy of Mulase and Rabin which appears more natural.  The
extension of the results of this paper to the supersymmetric case
will appear somewhere else.

\ack

We thank Sonia Stanciu for reading and commenting on a previous
version of this manuscript.  We also take this opportunity to thank
Boris Khesin for his interest and for sending us \[KZ].  In addition,
JMF is grateful to the Instituut voor Theoretische Fysica of the
Universiteit Leuven for its hospitality and support during the start
of this collaboration.
\refsout
\bye